\documentstyle[epsf,emulateapj]{article}

\newcommand{\etal}{{et al.}\hspace{1mm}}
\newcommand{\sbunits}{erg cm$^{-2}$ s$^{-1}$ arcmin$^{-2}$}
\newcommand{\sbspunits}{keV cm$^{-2}$ s$^{-1}$ sr$^{-1}$ keV$^{-1}$}
\newcommand{\hmpc}{$h^{-1}$ Mpc}

\begin{document}


\title{The X-ray surface brightness distribution from diffuse gas}

\author{Greg L. Bryan\altaffilmark{1}}
\affil{Department of Physics, Massachusetts Institute of Technology,
Cambridge, MA 02139}

\and

\author{G. Mark Voit}
\affil{Space Telescope Science Institute, 3700 San Martin Drive,
Baltimore, MD 21218}


\altaffiltext{1}{Hubble Fellow}


\begin{abstract}

We use simulations to predict the X-ray surface brightness
distribution arising from hot, cosmologically distributed diffuse gas.
The distribution is computed for two bands: 0.5-2 keV and 0.1-0.4 keV,
using a cosmological-constant dominated cosmology that fits many other
observations.  We examine a number of numerical issues such as
resolution, simulation volume and pixel size and show that the
predicted mean background is sensitive to resolution such that higher
resolution systematically increases the mean predicted background.
Although this means that we can compute only lower bounds to the
predicted level, these bounds are already quite restrictive.  Since
the observed extra-galactic X-ray background is mostly accounted for
by compact sources, the amount of the observed background attributable
to diffuse gas is tightly constrained.  We show that without physical
processes in addition to those included in the simulations (such as
radiative cooling or non-gravitational heating), both bands exceed
observational limits.  In order to examine the effect of
non-gravitational heating we explore a simple modeling of energy
injection and show that substantial amounts of heating are required
(i.e. 5 keV per particle when averaged over all baryons).  Finally, we
also compute the distribution of surface brightness on the sky and
show that it has a well-resolved characteristic shape.  This shape is
substantially modified by non-gravitational heating and can be used as
a probe of such energy injection.

\end{abstract}



\section{Introduction}

The X-ray background has now been largely resolved into individual
point sources (e.g. Hasinger \etal 1993; Mushotzky \etal 2000), the
majority of which are thought to be AGN (e.g. Boyle 1994).  This
constrains the amount of the background which may be due to a diffuse
hot intergalactic medium (Barcons, Fabian \& Rees 1991).  Recent work
has applied these ideas specifically to the soft X-ray background,
arguing that currently popular cosmological models predict too much
flux and so require some non-gravitational heating to reduce the
emissivity (Pen 1999; Wu, Fabian \& Nulsen 2000b).  However, since the
emission process (primarily Bremsstrahlung) is sensitive to
temperature and particularly density, it is necessary to have a good
prediction of the density and temperature distribution of the diffuse
gas.  To date, most work on this subject has adopted semi-analytic
techniques or made strong assumptions about the temperature
distribution of the gas.  It has lately become possible (although as
we will show still difficult) to use numerical simulations to model
the distribution of gas in the universe and so directly predict its
X-ray emissivity.

Beyond a simple number --- the mean background --- it is of interest
to predict other facets of the diffuse X-ray background, such as its
distribution function.  This last quantity tells us how much of the
sky has a given surface brightness and can be used to investigate the
superposition of multiple sources along the line of sight during
cluster or group observations, or during searches for filaments (Voit,
Evrard \& Bryan 2000).  It is also a probe of the physics of galaxy
formation, as its shape is sensitive to non-gravitational heating
from, for example, supernovae and AGN.  A number of groups have
examined the spatial correlation of the X-ray background with other
objects (e.g. Soltan \etal 1999), and although we do not do so here,
it would also be of interest to investigate the predicted clustering
properties of the diffuse X-ray background (e.g. Croft \etal 2000).

In this paper, we draw on the results of hydrodynamic simulations to
compute the distribution of surface brightness in two bands on the
X-ray sky, neglecting the contribution from point sources.  Since the
simulation boxes are relatively small compared to the line-of-sight
distance, we generate the distribution by effectively stacking the
simulations.  The method for doing this is described in
section~\ref{sec:calc}.  Using this algorithm, we show, in
section~\ref{sec:results}, the results from a number of simulations,
examing the effect of resolution and comparing to previous work on
this subject.  Finally, we investigate the effect of additional
physical processes, such as cooling and non-gravitational heating (from,
for example, supernovae).

In a previous paper (Voit, Evrard \& Bryan 2000), we have discussed
how the diffuse X-ray background can be a source of confusion for
group observations as well as during searches for filamentary gas.  In
a related paper (Voit \& Bryan 2000), we discuss these distributions
in the context of a simple analytic model which accurately reproduces
the simulation results, as well as examining the role point sources
play in the distribution.


\section{Computing the X-ray background}
\label{sec:calc}

In order to compute the X-ray background from the diffuse gas, we must
know the distribution of density and temperature along the
line-of-sight.  Rather than compute this in a single step, we break
down the computation into two steps.  In the first, we simulate a
cubic region of the universe, using a numerical hydrodynamics code
(described below).  The results of these simulations are saved at
various points during the computation, and each output can be used to
generate a map of X-ray emission from that particular region, at that
redshift.  In the second step, we develop an algorithm to
statistically combine these maps (or rather, the surface brightness
distributions computed from them) to generate the final distribution.

\subsection{The Simulations}

The density and temperature distribution is computed by solving the
equations of hydrodynamics in a comoving volume of side length $L$.
An adaptive mesh refinement (AMR) technique is used to solve these
equations.  This algorithm is described elsewhere (Bryan 1999; Bryan
\& Norman 1997, 1999; Norman \& Bryan 1998), but we briefly summarize
it here.  The dark matter is simulated by following particle
trajectories, while the baryon fluid is modeled by discretizing the
density, temperature and velocity distributions onto a mesh.  The
discretized equations of hydrodynamics are solved using a piece-wise
parabolic method, modified for cosmology (Colella \& Woodward 1984;
Bryan \etal 1995).  The gravitational acceleration comes from solving
Poisson's equation on the mesh, using an iterated multi-grid technique.

As objects collapse and form, the code must resolve smaller and
smaller length scales.  This is accomplished in AMR by overlaying
additional, finer meshes onto areas that require improved resolution
as the simulation proceeds.  These finer meshes have a cell spacing
1/2 as large as the coarser grids from which they obtain their
boundary conditions.  This procedure can be repeated recursively, with
finer and finer meshes covering less and less volume.  The cell
spacing on any given level $l$ (where $l=0$ refers to the top grid) is
given by $\Delta x$ = $L/(2^lN)$, where $N$ is the number of mesh
points per dimension on the top grid.  The refinement criterion is
designed to keep a fixed mass resolution: additional grids are added
whenever the mass (either baryonic or dark) in a cell exceeds a
certain threshold, chosen to be four times the initial mass in a cell.

The simulation is initialized at high redshift ($z = 30$), when the
perturbations are nearly linear.  In this work, we restrict ourselves
to a single cosmological model, one which matches a large number of
current observations.  The ratio of the density in non-relativistic
matter to the critical density is taken to be $\Omega_0 = 0.3$.  The
model is flat, with a cosmological constant energy density
$\Omega_{\Lambda} = 0.7$ and Hubble constant $h=0.67$, where $h$ is in
units of 100 km/s/Mpc.  The baryon fraction was taken to be $\Omega_b
= 0.04$, which is slightly on the low side of current estimates
(Burles \& Tytler 1999).  The power spectrum of initial density
perturbations is taken from Eisenstein \& Hu (1998), which we
normalize so that the {\it rms} fluctuations in spheres of 8
$h^{-1}$Mpc is $\sigma_8 = 0.9$.

In this paper, we will analyze the results from a number of
simulations with varying resolutions and box sizes in order to
investigate numerical uncertainties.  The parameters for these runs
are shown in Table~\ref{table:sims}, along with designated name for
each simulation.  In each case, we list the box size ($L$), the number
of grid point per dimension on the top grid ($N_{grid}$), the highest
resolution reached in the mesh refinement (in each case we go down
four levels).  Finally, the last two simulations include a very simple
heating scenario designed to investigate the effect of feedback.  In
this run, we increased the gas temperature by 1 keV at $z = 3$ (an
increase of 1.5 keV per particle).  Although this is obviously overly
simplified, it is a straightforward way to show the effect of
increasing the gas entropy on the X-ray background.

\begin{deluxetable}{cccccccc}
\footnotesize
\tablecaption{Simulations analyzed in this paper
\label{table:sims}}
\tablewidth{0pt}
\tablehead{
\colhead{designation} &
\colhead{$L (h^{-1} Mpc)$} &
\colhead{$N_{grid}$} &
\colhead{$M_{dm} (M_{\odot})$} &
\colhead{finest $\Delta x (h^{-1} kpc)$} &
\colhead{feedback} &
\colhead{$\bar{S}_{0.5-2.0}$} &
\colhead{$\bar{S}_{0.1-0.4}$}
}
\startdata
L100  & 100 & 128 & $7.9 \times 10^{10}$ & 49  & no &
 $1.5 \times 10^{-15}$ & $1.2 \times 10^{-15}$ \\
L50-  & 50  & 64  & $7.9 \times 10^{10}$ & 49  & no &
 $1.4 \times 10^{-15}$ & $1.2 \times 10^{-15}$ \\
L50+  & 50  & 128 & $9.9 \times 10^{9}$  & 24  & no &
 $2.5 \times 10^{-15}$ & $2.9 \times 10^{-15}$  \\
L20-  & 20  & 64  & $5.1 \times 10^{9}$  & 19.2 & no &
 $3.3 \times 10^{-15}$ & $3.8 \times 10^{-15}$  \\
L20+  & 20  & 128 & $6.3 \times 10^{8}$  & 9.6 & no &
 $5.0 \times 10^{-15}$ & $6.8 \times 10^{-15}$ \\
L50f- & 50  & 64  & $7.9 \times 10^{10}$ & 49  & yes &
 $4.3 \times 10^{-16}$ & $3.7 \times 10^{-16}$  \\
L50f+ & 50  & 128 & $9.9 \times 10^{9}$ & 24  & yes &
 $6.8 \times 10^{-16}$ & $7.5 \times 10^{-16}$  \\
\enddata
\tablecomments{$L$ is the simulation box size, $N_{grid}$ is the
number of points per dimension in the initial grid, $M_{dm}$ is the
dark matter particle mass, $\Delta_x$ is the smallest cell size and
$\bar{S}$ is the mean predicted surface brightness in units of \sbunits.}
\end{deluxetable}


\subsection{Computing the distribution}

We compute the surface brightness distribution in the following way.
First, for each simulation output at a redshift $z$, we compute a
surface brightness map by integrating along lines-of-sight from one
edge of the volume to the other:
\begin{equation}
S(z, \Delta z) = \frac{1}{4 \pi (1+z)^4} \int \epsilon(T) n_e n_H dl,
\end{equation}
where $\epsilon(T)$ is the emissivity in the (appropriately
red-shifted) X-ray band of interest, $n_e$ is the electron density and
$n_H$ is the proton density.  The electron and proton densities are
computed assuming complete ionization, which is an excellent
approximation for the hot gas producing the emission.  The emissivity
is computed with a Raymond-Smith (1977; 1992 version) code assuming a
constant metallicity of 0.3 solar (as we will show below, most of the
background comes from groups and clusters at moderate redshifts so
this is a very reasonable approximation).  An example of such a map is
shown in Figure~\ref{fig:map1}.

\vspace{\baselineskip}
\epsfxsize=3in 
\centerline{\epsfbox{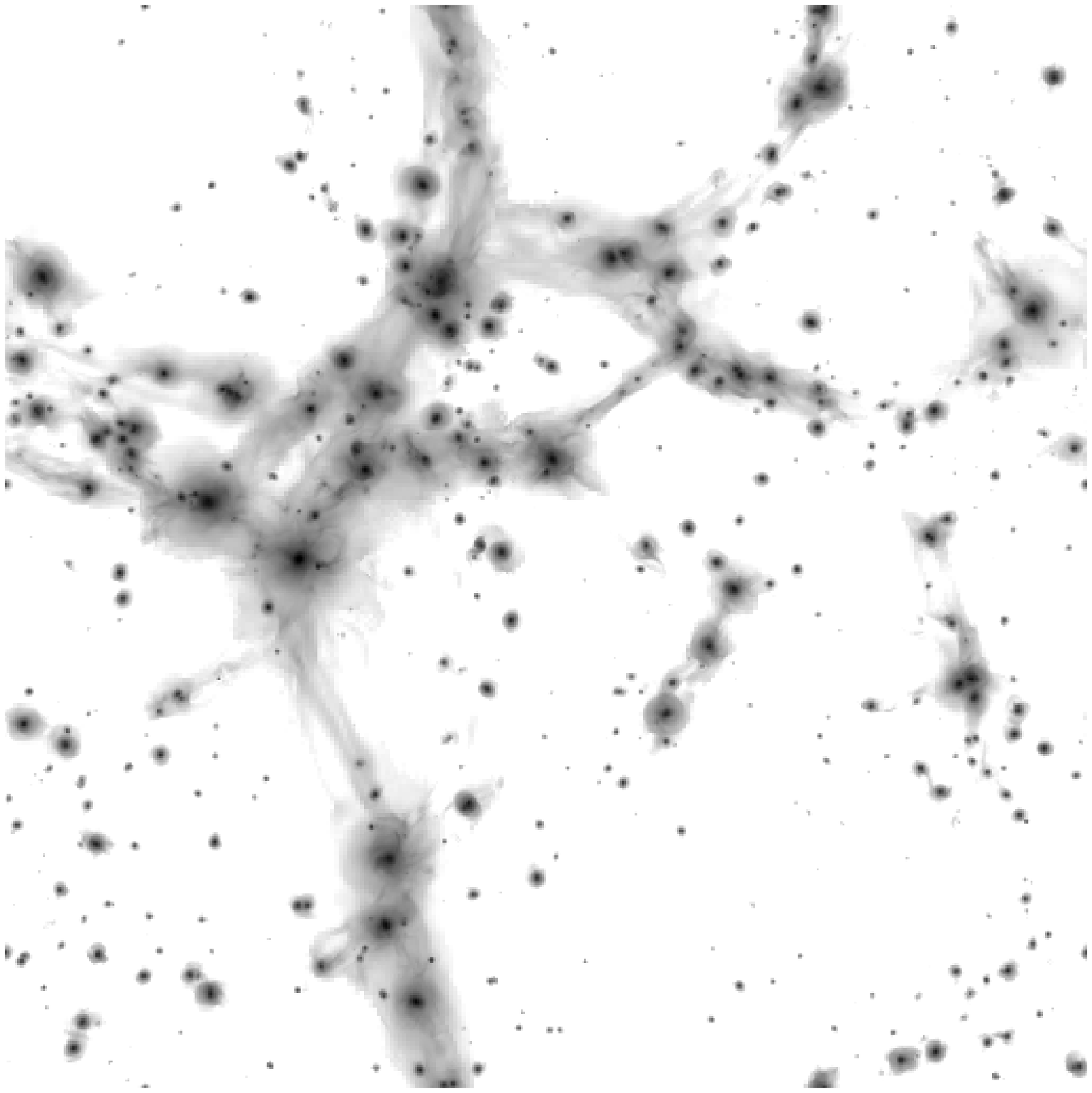}}
\figcaption{
A simulated 0.5-2 keV X-ray surface brightness map from a region 50
$h^{-1}$ Mpc on a side at redshift $z=0.4$ and line-of-sight distance
$\Delta z = 0.02$. The greyscale is logarithmic in order to bring out
the low surface brightness filaments and ranges from $10^{-21}$ to
$10^{-15}$ \sbunits.  The image is 178 arcmins on a side.
Note that this is the background from diffuse
sources only. 
\label{fig:map1}}
\vspace{\baselineskip}

This map is obviously not a complete X-ray image since it only
corresponds to the redshift range from $z-\Delta z/2$ to $z+ \Delta
z/2$, where $\Delta z = L/3000 E(z)$ and $L$ is the length of the
computational volume in comoving $h^{-1}$Mpc (this is accurate as long
as $\Delta z/z \ll 1$).  The cosmological term is $E^2(z) = \Omega
(1+z)^3 + \Omega_R (1+z)^2 + \Omega_\Lambda$ (the curvature component
is defined as $\Omega_R = 1 - \Omega - \Omega_\Lambda$; Peebles 1993).
Each pixel subtends an angle on the sky of $\theta = \Delta x / D_A$
where $\Delta x$ is the proper distance between pixels and $D_A$ is
the angular diameter distance.  We can also smooth the image at this
point, to correspond to an instrument with a given resolution.  We
discuss this point in more detail below; however, for our standard
computation, we do not smooth the maps.

From maps such as the one shown in Figure~\ref{fig:map1}, we can
compute both the mean surface brightness, $\bar{S}(z, \Delta z)$, but
also the full distribution function $dP/dS(S, z, \Delta z)$.  Here,
$P(S)$ is the probability that a given line-of-sight will have a
surface brightness less than $S$ (excluding compact sources).  This
distribution is normalized such that:
\begin{equation}
\int \frac{dP(S, z, \Delta z)}{dS} dS = 1.
\end{equation}
In practice, this is computed for 100 logarithmically distributed
points from $10^{-21}$ to $10^{-13}$ \sbunits so that the distribution
we actually generate is $dP/d\ln S (S, z, \Delta z)$.  The number of
redshifts ($z_i$) for which we compute distributions varies somewhat,
depending on the details of the simulation, but is typically about 20.
In Figure~\ref{fig:dpds_0.5_2.0}, we show the individual $dP/dS(S, z,
\Delta z)$ for a range of redshifts.

It is relatively easy to compute the mean predicted background,
including contributions from all redshifts; this is given by:
\begin{equation}
\bar{S} = \int \frac{d\bar{S}(z)}{dz} dz
\end{equation}
where the mean surface brightness per unit redshift at a given
arbitrary redshift $z$ is linearly interpolated from the tabulated
redshifts $z_i$ (i.e. $d\bar{S}_i(z)/dz = \bar{S}_i(z, \Delta
z)/\Delta z$).  The results are given in Table~\ref{table:sims}.

The differential distribution is more difficult to compute, however it
can be built out of the individual distribution functions.  First, we
examine the problem of combining two distributions at different
redshifts, $z_1$ and $z_2$.  If we make the simplification that the
spatial correlations on scales larger than the box size are
negligible, then the joint distribution is given by:
\begin{equation}
\frac{dP_J(S)}{dS} = \int_0^S
       \frac{dP(S^\prime, z_1, \Delta z_1)}{dS}
       \frac{dP(S - S^\prime, z_2, \Delta z_2)}{dS} dS^\prime.
\label{eq:joint}
\end{equation}
When performing this numerical integration, some care must be taken
due to the logarithmic spacing of the functions.  We divide the sum
into two parts, split at $S/2$.  In the first half ($S^\prime <
S/2$), we evaluate the function at the points at which
$dP(S^\prime, z_1, \Delta z_1)/dS$ was determined.  In the second half
($S^\prime > S/2$), the points of the function $dP(S-S^\prime, z_2,
\Delta z_2)/dS$ are used.  This insures that the finest spacing
available is used at all times.

This combines two distributions, but it is a natural extension to
combine any number, since we can simply re-apply Eq.~(\ref{eq:joint})
with the first term replaced by the joint distribution $dP_J(S)/dS$
and the second by a distribution from another redshift $z_3$.  The
result will be referred to as the cumulative distribution
$dP_C(S)/dS$.

To systematically stack the simulations, we adopt the following
procedure.  Starting at redshift $z = 10 - \Delta z_0 / 2$ (the
contribution from larger redshifts is very small, see
Figure~\ref{fig:dSdz}), we set the cumulative distribution to
be equal to $dP(S, z_0, \Delta z_0)/dS$.  Ideally, we would like to
simply be able to stack the simulation boxes so that one end matches
the next; however, in general this requires many outputs and, for a
statistical determination of the distribution function, it is not
necessary.  Instead, we take a step in redshift of size $\Delta z(z)$,
where $\Delta z(z)$ is linearly interpolated from the outputs at $z_i$
and $z_{i+1}$ which bracket $z$ (i.e. $z_i < z < z_{i+1}$).  We
convolve the current distribution $dP_C(S)/dS$ with a distribution
which is similarly interpolated from the distributions computed at
$z_i$ and $z_{i+1}$.  The current redshift, $z$, is then decreased by
the amount $\Delta z(z)$ and the procedure continues.  We stop at $z
\approx 0.1$, at which point the pixels become very large (and
individual sources would be clearly apparent in the X-ray sky).  This
introduces a small uncertainty in the final background.

In Figure~\ref{fig:dpds_0.5_2.0}, we show the resulting distribution
from the L100 simulation, for the 0.5 to 2.0 keV band.  We plot it in
two different ways.  The first (top panel) shows $dP/d\ln S$, the
probability that a given pixel will fall in a given logarithmic
interval in surface brightness.  Thus we see that most pixels would
have a surface brightness of a few $\times 10^{-16}$ \sbunits (from
diffuse emission alone).  The second plot shows this quantity weighted
by another factor of $S$ in order to show the quantity that is
important for computing the mean flux.  We see that most of the
contribution to the mean comes from pixels with a surface brightness
of $\sim 10^{-14}$ \sbunits.

\vspace{\baselineskip}
\epsfxsize=3.5in
\centerline{\epsfbox{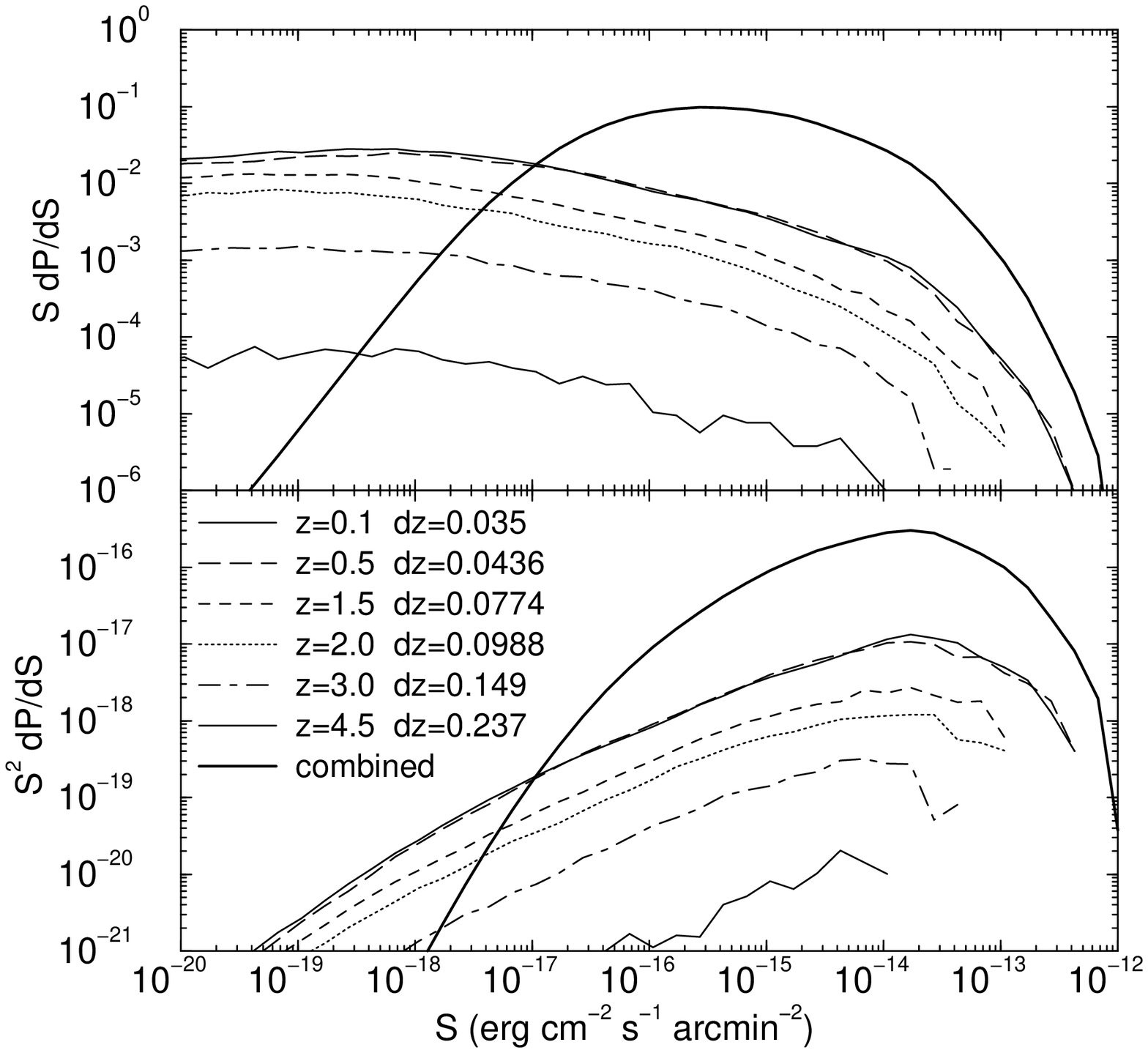}}
\figcaption{The X-ray surface-brightness distribution in the 0.5-2.0 keV
band from the L100 simulation.  The dashed lines show the
contributions from various simulation boxes at the redshifts
indicated, while the bold line is the final, composite distribution.
The top panel plots $S dP/dS$ in order 
to show the relative number of pixels per logarithmic interval in
surface brightness.  The bottom distribution multiplies this by
another factor of $S$ to show where the contribution to the mean
surface brightness originates.
\label{fig:dpds_0.5_2.0}}
\vspace{\baselineskip}

Comparing the individual distributions (from short redshift ranges) to
the cumulative distribution, we can see the effect of stacking.  For
low surface brightness pixels, the distribution is strongly decreased.
This arises simply from the fact that while for a short line of sight
(of order $\Delta z \sim 0.03$), it is possible to miss all of the
dense knots and filamentary structures seen in Figure~\ref{fig:map1},
it is extremely unlikely for a full line-of-sight.  On the high end of
the distribution, the shape is preserved while the amplitude
increases.  This is due to rare, bright pixels for which a single
passage through a massive cluster or group dominates the surface
brightness contribution.  Since multiple passages are rare, increasing
the length of the line-of-sight just increases the number of such
pixels; their chance of overlap is negligible.

We can double-check the calculation of the distribution by using it to
compute the mean, since:
\begin{equation}
\bar{S} = \int S \frac{dP}{dS} dS / \int \frac{dP}{dS} dS
\end{equation}
Since the distribution and the mean are computed independently (as
described above), these two estimates for $\bar{S}$ should agree.
For the larger box (50 and 100$h^{-1}$ Mpc), the agreement is very
good, within 10\%.  For the smallest box, the difference is larger,
around 20\%, as might be expected from the very large number of
redshift steps that have to be taken for the smallest simulated box.

In Figure~\ref{fig:dSdz}, we show how the contribution to the mean
flux depends on redshift.  Notice that the differential contribution
($d\bar{S}/dz$) is a monotonic function of redshift, implying that
most of the contribution comes from relatively low redshifts.

\vspace{\baselineskip}
\epsfxsize=3.3in
\centerline{\epsfbox{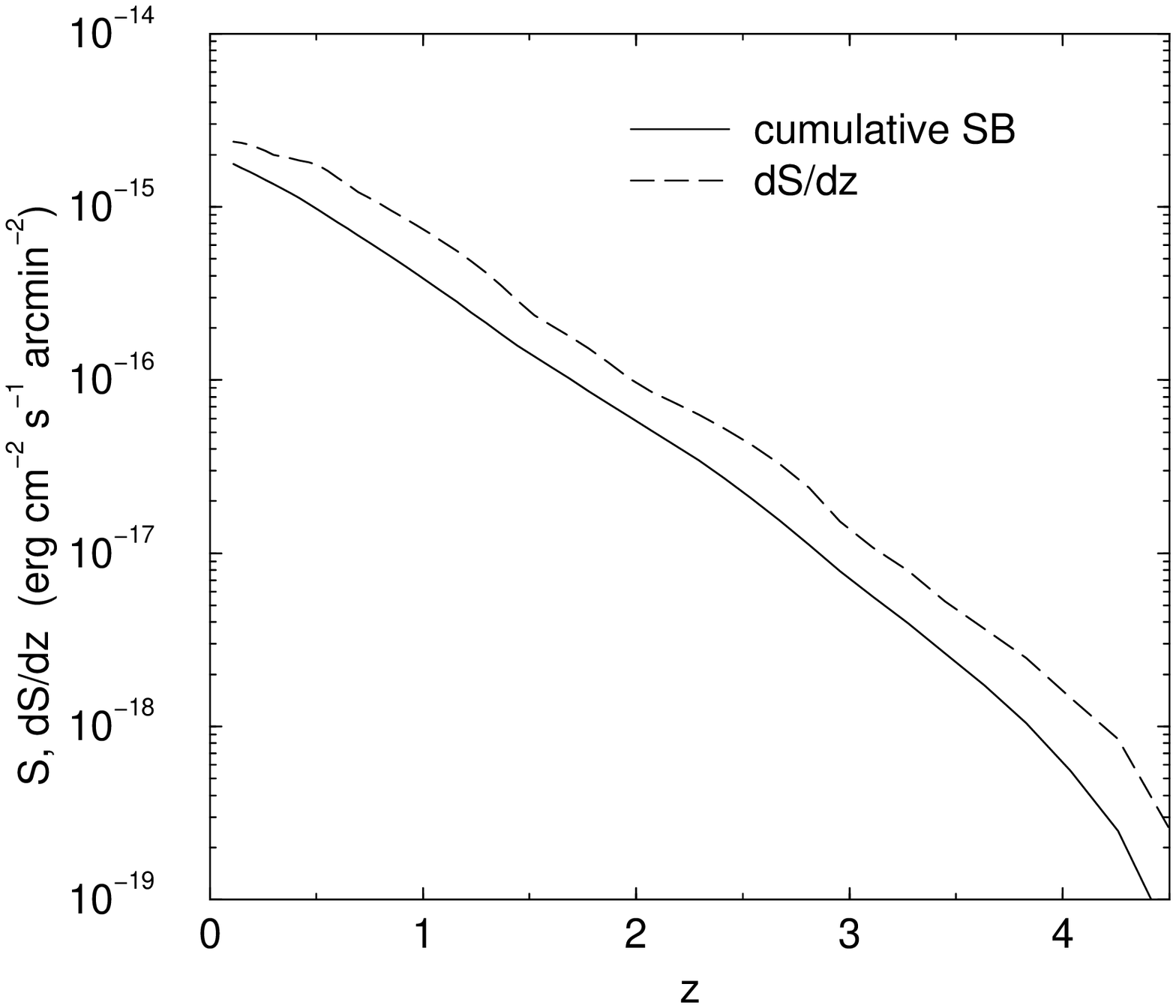}}
\figcaption{This plot shows the mean background diffuse flux as a
function of redshift in the 0.5 to 2.0 keV band.  The solid line is
the cumulative flux (i.e. $\bar{S}(z>z_0)$, while the dashed is the
differential contribution ($d\bar{S}/dz$).
\label{fig:dSdz}}
\vspace{\baselineskip}


\section{Results}
\label{sec:results}

The mean fluxes for the no-cooling, no-feedback simulations are larger
than observed.  However, before we draw conclusions from this, we
first examine some numerical issues in order to get some idea of the
robustness of the result.

\subsection{Numerical Issues}
\label{sec:numerical}

The X-ray emitting diffuse gas tends to be distributed on large
scales, but because of its density-squared emissivity law, depends
sensitively on the small scale distribution within groups, clusters
and filaments.  This implies that both box size and resolution will be
issues.  Here we describe a resolution study designed to examine these
effects.  The different simulation box sizes and resolutions are shown
in Table~\ref{table:sims}, along with the resulting mean background
computed using the methods described earlier.

Remarkably, it appears that the results are more sensitive to the
resolution of a given simulation, and nearly insensitive to the size
of the box (at least for the range considered here).  For example, the
L100 and L50- simulations differ substantially in the amount of large
scale structure which is captured in the simulation, but have the same
spatial and mass resolution.  Their computed mean background fluxes
are nearly identical.

The mass resolution of the simulation appears to be the most important
factor.  For the background fluxes in the 0.5-2.0 keV band from
Table~\ref{table:sims} (excluding those with pre-heating), the results
can be described by the fitting function:
\begin{equation}
\bar{S} = 2.3 \times 10^{-15} (M_{dm}/10^{10}
M_{\odot})^{-0.24} \hbox{\sbunits}.
\end{equation}
The correlation coefficient for this fit is very high, 0.996, leaving
very little room for other effects such as box size (we note that for
even harder bands than considered here, such as 2-10 keV, a larger
volume may be important).  A naive extrapolation of this trend to
infinite resolution predicts an infinite contribution to the soft
X-ray background.  In reality, the contribution levels off as all the
structures that are hot enough to emit in this band are resolved.  In
fact, it seems plausible that we are very close to this regime, since
the virial temperature of a $10^{13}$ M$_{\odot}$ object is about 0.1
keV, which is about the lowest temperature which can contribute
significantly to the harder band .  At the highest resolution, such
objects are resolved with $\sim 10^4$ particles, which experience
indicates is the minimum necessary to resolve the central regions
where most of the emission originates.  See also Bryan \& Norman
(1998) for a discussion of how numerical resolution affects the
predicted X-ray luminosity of clusters.  Given this discussion, we can
quote only a lower limit to the predicted soft X-ray background from
diffuse gas (without pre-heating or radiative cooling):
\begin{equation}
\bar{S}_{0.5-2.0} \ge 5.0 (\Omega_b h^2/0.018)^2 \times 10^{-15}
\hbox{\sbunits}
\label{eq:slim_0.5_2.0}
\end{equation}

\vspace{\baselineskip}
\epsfxsize=3.3in
\centerline{\epsfbox{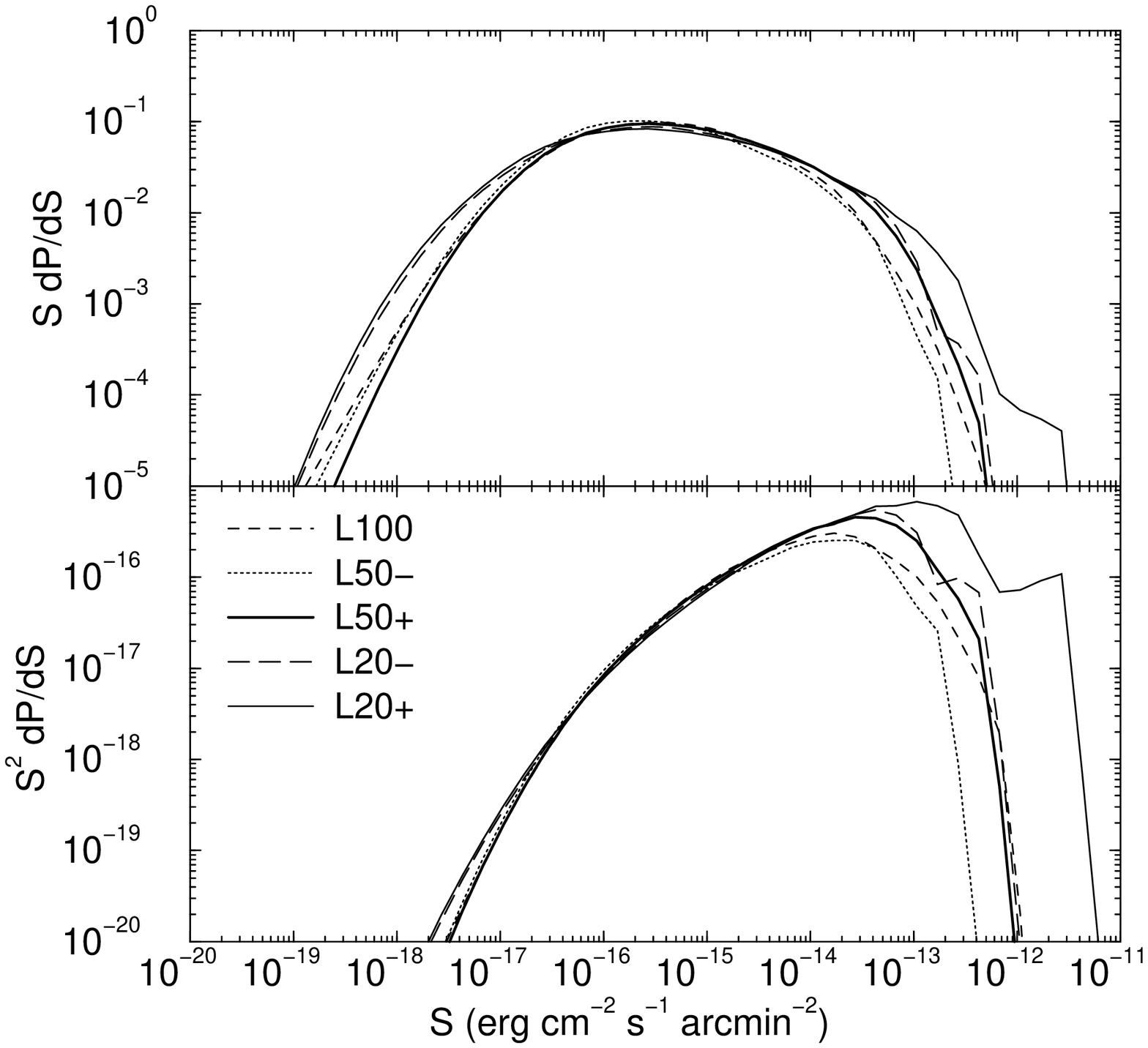}}
\figcaption{The distribution function in the 0.5-2.0 keV band for a
  variety of simulations with different resolutions.  The top and
  bottom panels have the same meaning as in
  Figure~\ref{fig:dpds_0.5_2.0}.
\label{fig:dpds_resol}}
\vspace{\baselineskip}

In Figure~\ref{fig:dpds_resol}, we show the computed distribution
functions for some of the simulations described in
Table~\ref{table:sims} in the 0.5-2.0 keV band.  The most striking
feature of this plot is the gross similarity amoung the curves; this
is quite reassuring given the very different resolutions and box sizes
used.  Closer examination reveals a number of systematic trends.
First, the range from $10^{-17}$ to $10^{-14}$ \sbunits is quite
robust.  Below this, in the low surface brightness domain, the two
larger box sizes produce nearly identical results irrespective of
resolution, while the $L=20$ \hmpc~simulations show an elevated
distribution function (for both resolutions).  This demonstrates that
the minimum box size to obtain an un-biased sample of the distribution
is between 20 and 50 \hmpc.

For the high surface brightness end, it's clear that resolution plays
an important role, with high-resolution simulations producing a larger
number of high-brightness pixels.  In fact, as we will see, these
pixels are primarily produced in the centers of groups and clusters.
The bottom panel of Figure~\ref{fig:dpds_resol} shows that it is this
high end which determines the mean flux, causing the resolution
dependence discussed earlier.  The kink at large S values in the
L20- curve results from a single cluster, indicating the difficulty
in obtaining a fair sample in such a small volume (although note that
this single cluster does not make a dominant contribution to the mean
surface brightness).  Based on this discussion, we have a robust
determination of the diffuse background flux distribution below about
$10^{-13}$ \sbunits.  Remarkably, this accounts for some 99\% of the
sky, although the remaining $\sim 1\%$ of the pixels are responsible
for setting the mean background.  We remind the reader at this point
that these simulations include only a minimal physics model, and
exclude the effects of both radiative cooling and stellar feedback.

\vspace{\baselineskip}
\epsfxsize=3.5in
\centerline{\epsfbox{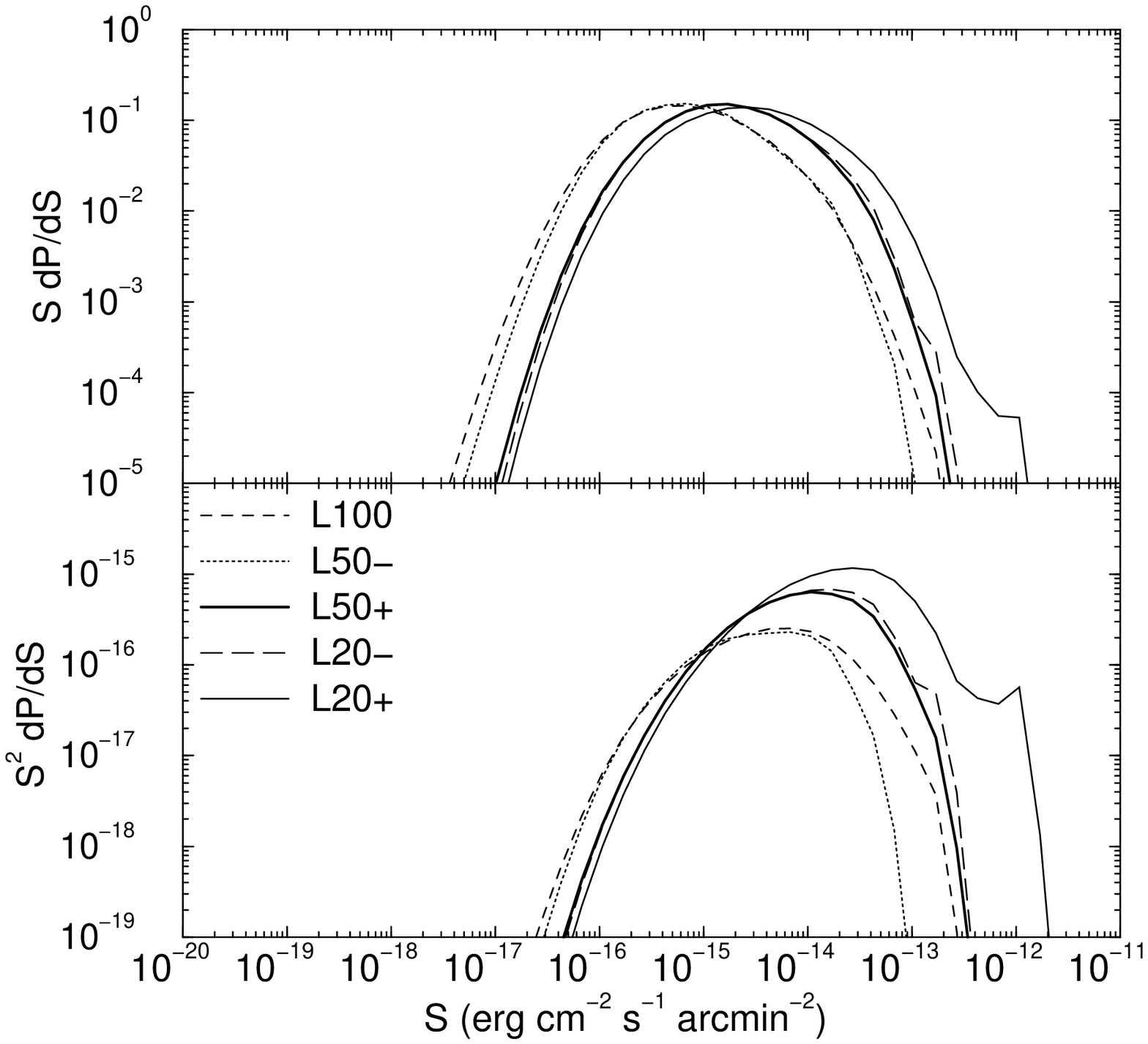}}
\figcaption{
The distribution in the 0.1-0.4 keV band function for a variety of
simulations with different resolutions.
\label{fig:dpds_resol_0.1_0.4}}
\vspace{\baselineskip}

We now turn to the softer, 0.1-0.4 keV, band.  The mean backgrounds
given in the last column of Table~\ref{table:sims} show the same
systematic behaviour with resolution as for the harder band, so again
we can estimate only a minimum contribution to the background from
diffuse gas (assuming no cooling or re-heating):
\begin{equation}
\bar{S}_{0.1-0.4} \ge 6.8 (\Omega_b h^2/0.018)^2 \times 10^{-15}
\hbox{\sbunits}
\label{eq:slim_0.1_0.4}
\end{equation}

The distribution functions shown in
Figure~\ref{fig:dpds_resol_0.1_0.4} are much more affected by
resolution than for the harder band.  In fact, as the resolution
increases, the entire curve shifts to higher surface brightness
levels.  The effect of box size is very small or negligible, even for
L=20 \hmpc.  The resolution effect occurs because the covering factor
for the smaller, more numerous objects which contribute to the soft
band is nearly unity.  Since resolution affects all of these objects
(which are small to begin with), the net result is a uniformly
brighter background.  The unfortunate result is that we cannot
robustly predict the background distribution in this band (although it
does appear that the shape itself is fairly robust).


\subsection{Smoothing the distribution}

So far we have focussed on the effect of the simulation resolution on
the distribution function of the diffuse X-ray background.  However,
there are also observational uncertainties which can systematically
bias the result.  One of these is the removal of point sources, which
we will not treat in this paper.  However, another which we can
address is the effect of artificial smoothing on the sky.  This can
occur either unintentionally, due to inherent limitations in the
resolution of the instrument (X-rays are certainly more difficult to
focus than optical photons).  However, even for such high-resolution
instruments as Chandra, there may be reasons to introduce smoothing as
a post processing step.  This might be done to improve the photon
statistics in areas of the sky with very low surface-brightness.  

Figure~\ref{fig:dpds_smooth} shows the effect of various amounts of
smoothing on the 0.5-2.0 keV distribution of the L50+ simulation.
In each case, we have smoothed the distribution with a Gaussian kernel
with widths ranging from 0 up to 180 arcsecs.  Predictably, the upper
and lower ends of the distribution are truncated and with 180''
smoothing the sky shows only gentle fluctuations.  This confirms the
visual impression in the maps shown earlier that the smallest features
(generally the peaks of some clusters and groups) are only a few
arcsecs in size.

\vspace{\baselineskip}
\epsfxsize=3.5in
\centerline{\epsfbox{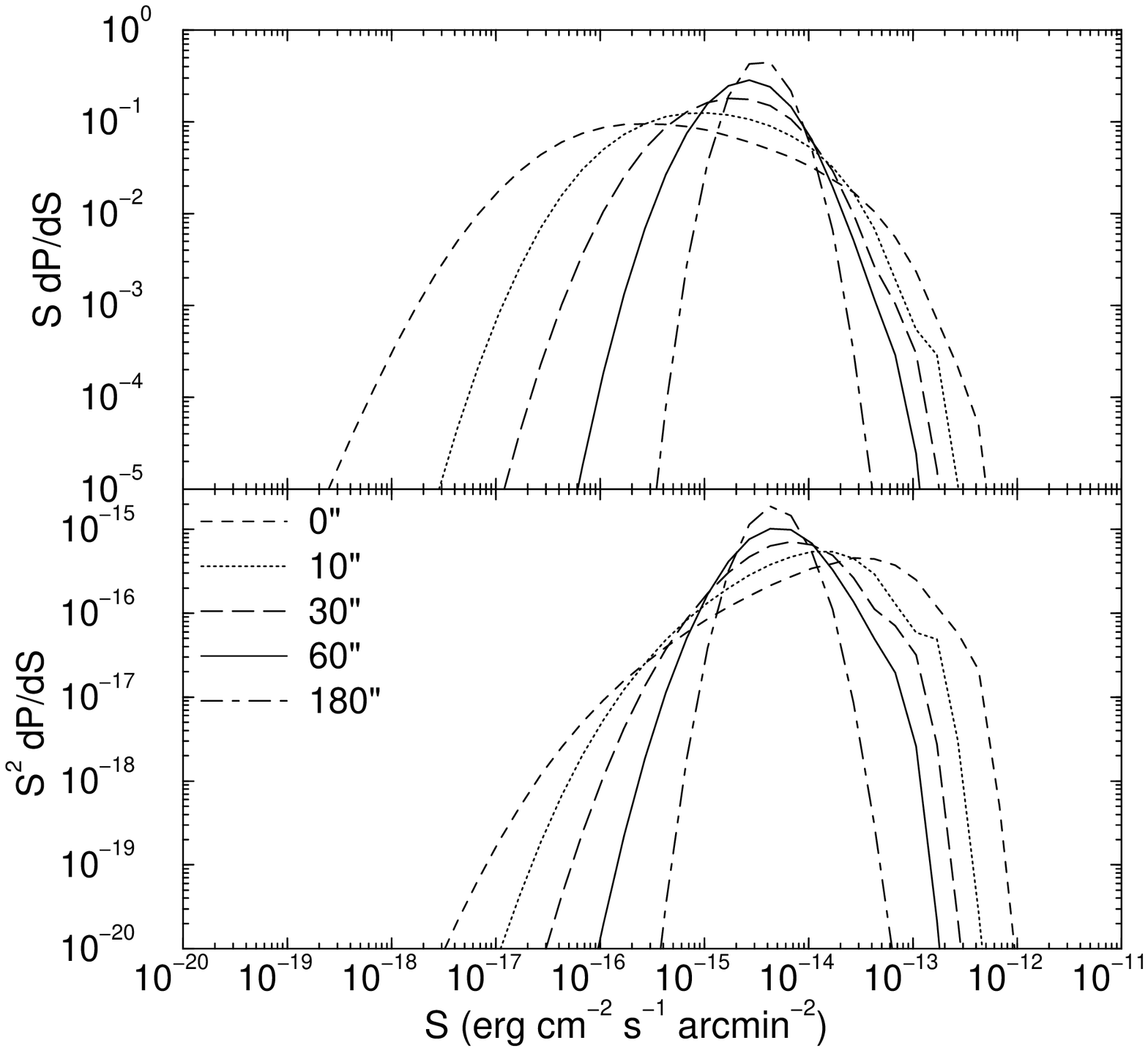}}
\figcaption{The distribution function for the L50+ simulation in the
0.5 to 2.0 keV band.  In computing the distribution function, we
have used a variety of Gaussian smoothing lengths, ranging from
no additional smoothing up to 3 arcminutes.
\label{fig:dpds_smooth}}
\vspace{\baselineskip}

We note that a single smoothing length is probably not the most
efficient use of the data.  A reasonable way to ensure constant
signal-to-noise for all measured surface-brightness values would be to
introduce an adaptive smoothing algorithm.  This would preserve the
high end of the distribution where the smoothing length would remain
small, while introducing the minimum amount of smoothing in the low
end.


\subsection{Cooling and non-gravitational heating}

There are two physical processes which we have so far neglected in
these simulations.  One of these is radiative cooling, which will have
two contradictory effects.  The first will be to increase the density
of some of the X-ray emitting gas, since gas which cools will tend to
be compressed.  This will enhance the emissivity and hence the diffuse
background.  On the other hand, if the gas cools sufficiently rapidly
that it never gets heated to X-ray emitting temperatures (or only
briefly to such temperatures), then the energy will be emitted at
other wavelengths, leading to a decrease in the diffuse X-ray
background.  To follow this process numerically is a computationally
demanding task, since the simulation must resolve scales from below a
kpc to 50 Mpc, accompanied with high mass resolution.  Due to this
difficulty, we are unable to investigate this process numerically.
We simply note that it remains a viable mechanism for decreasing the
X-ray background level to match observations (e.g. Bryan 2000; Croft
\etal 2000).

The other physical process is feedback from the stellar systems within
galaxies, primarily supernovae.  Although we cannot follow the
formation of galaxies and stars in detail for the same reasons
described above, we can investigate simple energetic prescriptions
which mimic the more complicated physics.  Perhaps the most simplified
way to account for the effect of feedback is to imagine it occurring
at a single epoch with uniform efficiency in terms of energy per
baryon.  Exactly this prescription (or slight modifications thereof)
has been modeled in a number of recent papers and seems to well
account for the observed slope of the luminosity-temperature relation
of groups and clusters of galaxies (Cavaliere, Menci \& Tozzi 1997;
Wu, Fabian \& Nulsen 2000a; Valageas \& Silk 1999; Loewenstein 2000).
Although there is some disagreement on exactly how much heating is
required, typical values are around 1 keV per baryon.  We repeat two
simulations but add 1.5 keV per particle (suddenly) at z=3.  We will
loosely refer to these as ``feedback'' simulations, and designate them
as L50f+ and L50f-.

Heating results in a substantial decrease in the predicted mean
diffuse X-ray background (see Table~\ref{table:sims}).  The effect on
the distribution function is extremely strong, as shown in
Figure~\ref{fig:dpds_feedback} for the 0.5 to 2.0 keV band.  Due to
the increase in temperature of the low density material in filaments,
there is enhanced emissivity (within the band) in these regions and
hence fewer low surface brightness pixels.  The input of energy in
dense regions results in a decrease in the density in those regions
(groups become ``puffier'').  This decreases the number of
high-brightness pixels.  Together these changes result in a much more
peaked distribution, so that most pixels have a surface brightness
around $10^{-16}$ \sbunits.  On the other hand, the distribution is
nearly flat in $S^2 dP/dS$ so that a wide range contributes to the
mean value of the diffuse background, $\bar{S}$.

In this figure, we also plot the results for two ``feedback''
simulations with different resolutions.  Clearly, the same
numerical effects are operating as before.  This is
reflected in the systematic increase in the predicted mean flux with
resolution seen in Table~\ref{table:sims}.

The effect of feedback is also very striking in the other band.
Again, the distribution is
strongly peaked, and the mean background decreases substantially.
Also as before, the effect of resolution is such that we can only
predict a minimum mean surface brightness from diffuse gas.  However,
as we will see in section~\ref{sec:observations}, even this lower
limit is quite restrictive.

The feedback prescription adopted here is quite simple --- 1.5 keV of
extra energy per baryon everywhere at $z=3$.  If the energy input in
the real universe is concentrated around galaxies, then it may be
biased towards generally high density regions.  This should not change
the behaviour on the high $S$ end of the distribution (or on the mean
flux), but may reduce the size of the effect at the low
surface-brightness end, blunting the highly peaked structure seen in
the upper panel of Figure~\ref{fig:dpds_feedback}.

\vspace{\baselineskip}
\epsfxsize=3.5in
\centerline{\epsfbox{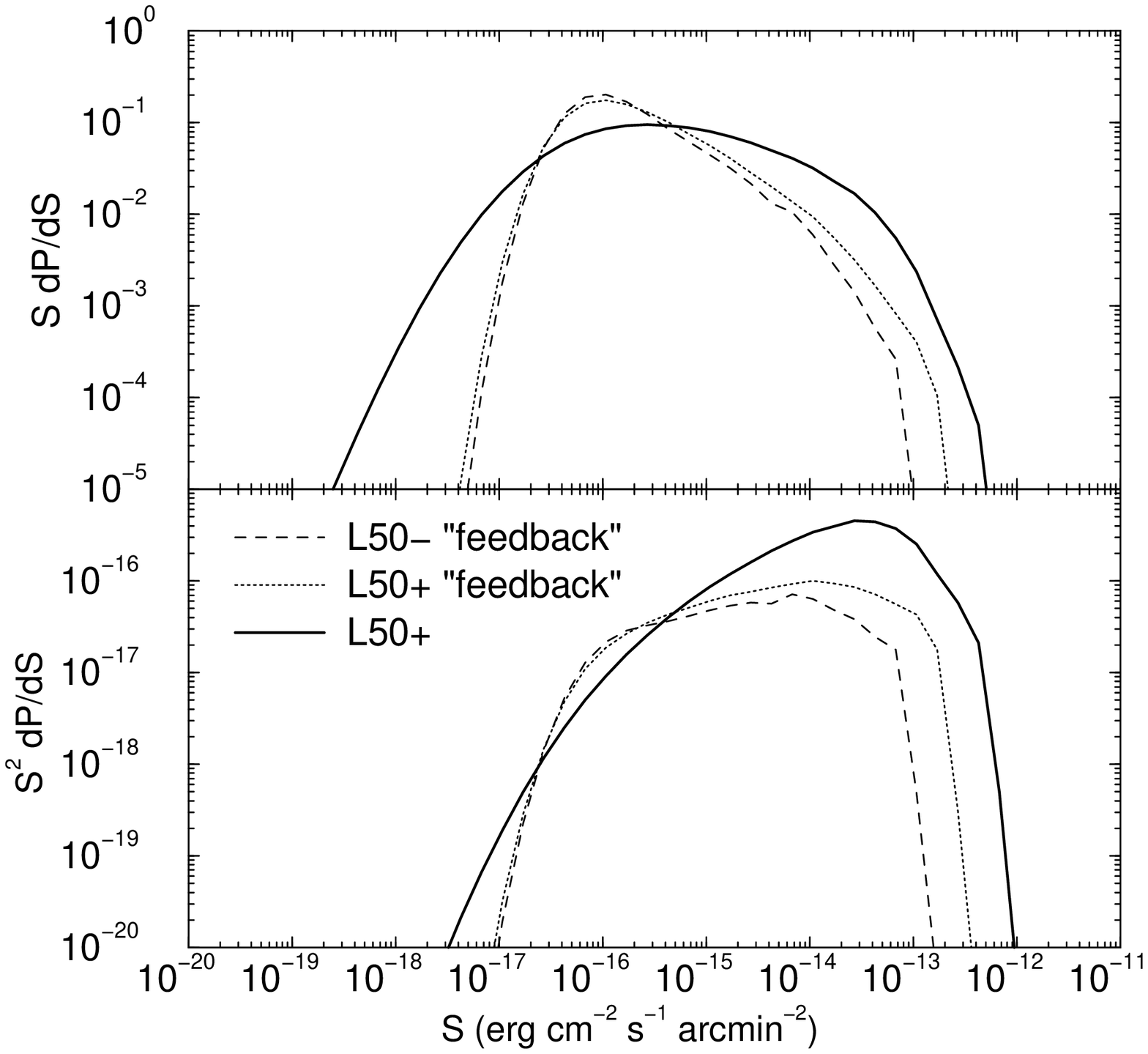}}
\figcaption{The 0.5-2.0 keV surface brightness distribution function
for the L50+ and L50- simulations (differing in resolution) with and
without the addition of 1.5 keV per baryon at z=3.  The top and bottom
panels have the same meaning as in Figure~\ref{fig:dpds_0.5_2.0}.
\label{fig:dpds_feedback}}
\vspace{\baselineskip}



\subsection{Where is the gas that emits the diffuse X-ray background?}

Figure~\ref{fig:map1} shows the surface brightness map from one of the
simulation volumes, and provides some hints as to where the emission
come from.  Clearly, the filaments which, at $z=0.4$, have typical
surface brightness values of around $10^{-16}$-$10^{-18}$ \sbunits
fall at the very low end of the expected distribution.  As discussed
in Voit, Evrard \& Bryan (2000), this makes them very difficult to see
due to confusion from other diffuse sources.  This is in addition to
the background from AGN and other point sources (see also Pierre,
Bryan \& Gastaud 2000).  On the other hand, the centers of the
clusters and groups provide the pixels with the highest
surface-brightness.

\vspace{\baselineskip}
\epsfxsize=2.8in
\centerline{\epsfbox{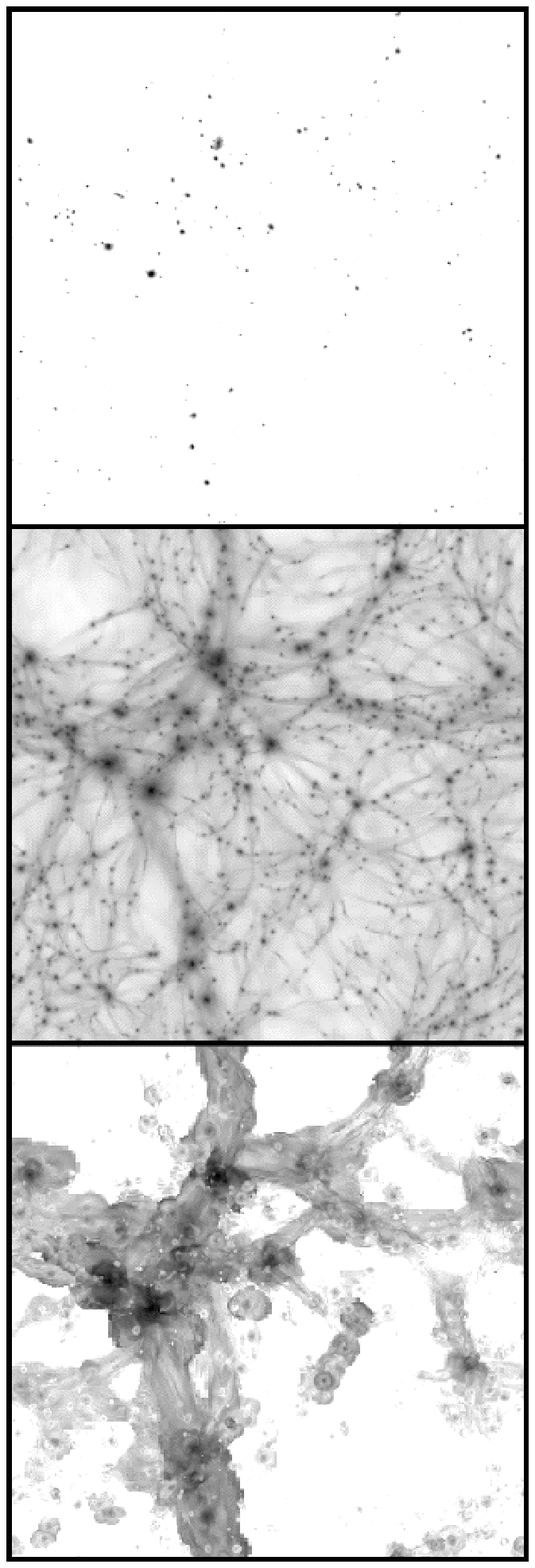}}
\figcaption{The top panel shows the 0.5-2.0 keV X-ray surface
brightness map from a region 50 $h^{-1}$ Mpc on a side at redshift
$z=0.4$ and line-of-sight distance $\Delta z = 0.02$.  The greyscale
is a stretch from $8.5 \times 10^{-16}$ to $8.5 \times 10^{-14}$
\sbunits, a range which is responsible for 75\% of the total flux.
For a larger stretch of the same region, showing filaments, see
Figure~\ref{fig:map1}.
The second frame shows a logarithmic greyscale map of the baryonic
density and the bottom shows the (emission-weighted) temperature map,
ranging from $10^6$ to $5 \times 10^7$ K.
\label{fig:sb_images}}
\vspace{\baselineskip}

While this gives a qualitative answer, it would be useful to know if
collapsed objects contribute the majority of the emission, and if so,
what size group or cluster is primarily responsible.  We answer these
questions in two ways.  The first is to pick out the surface
brightness levels that are responsible for most of the flux.  In
Figure~\ref{fig:sb_images}, we show the same simulated volume as in
Figure~\ref{fig:map1}, but now highlighting the regions for which the
surface brightness lies between $S = 8.5 \times 10^{-16}$ and $8.5
\times 10^{-14}$ \sbunits.  All together, these pixels account for
75\% of the total flux emitted from the entire volume.  Also shown is
a map of the projected baryonic density and the (emission-weighted)
temperature.  Clearly, most of the flux comes from collapsed regions
and not from filaments.  This is despite the fact that a large
majority of the gas is in filaments and diffuse structures.  This
implies that the gas distribution is iceberg-like: only a small
fraction of it is easily visible.

Having determined that most of the emission comes from collapsed
objects, it is of interest to determine what range of objects are
primarily responsible for the emission.  Although it is clear that
larger objects tend to have a higher surface brightness, they are also
rarer.  In Figure~\ref{fig:sb_tlum}, we show the relation between mean
surface brightness and luminosity-weighted temperature (closely
related to mass) for objects identified in the same simulation volume
discussed above.  The objects are identified in three-dimensions with
the hop halo finder algorithm (Eisenstein \& Hut 1999).  The virial
mass ($M_{200}$) is defined as all the mass within a sphere of radius
$r_{200}$, within which the mean density is 200 times the critical
density.  The mean surface brightness $S$ within the virial radius of
the object ($r_{200}$) is given by $L/(4 \pi^2 r_{200}^2 (1+z)^4)$ and
$L$ is the luminosity in the energy band of interest (defined in the
observer's frame).

\vspace{\baselineskip}
\epsfxsize=3.5in
\centerline{\epsfbox{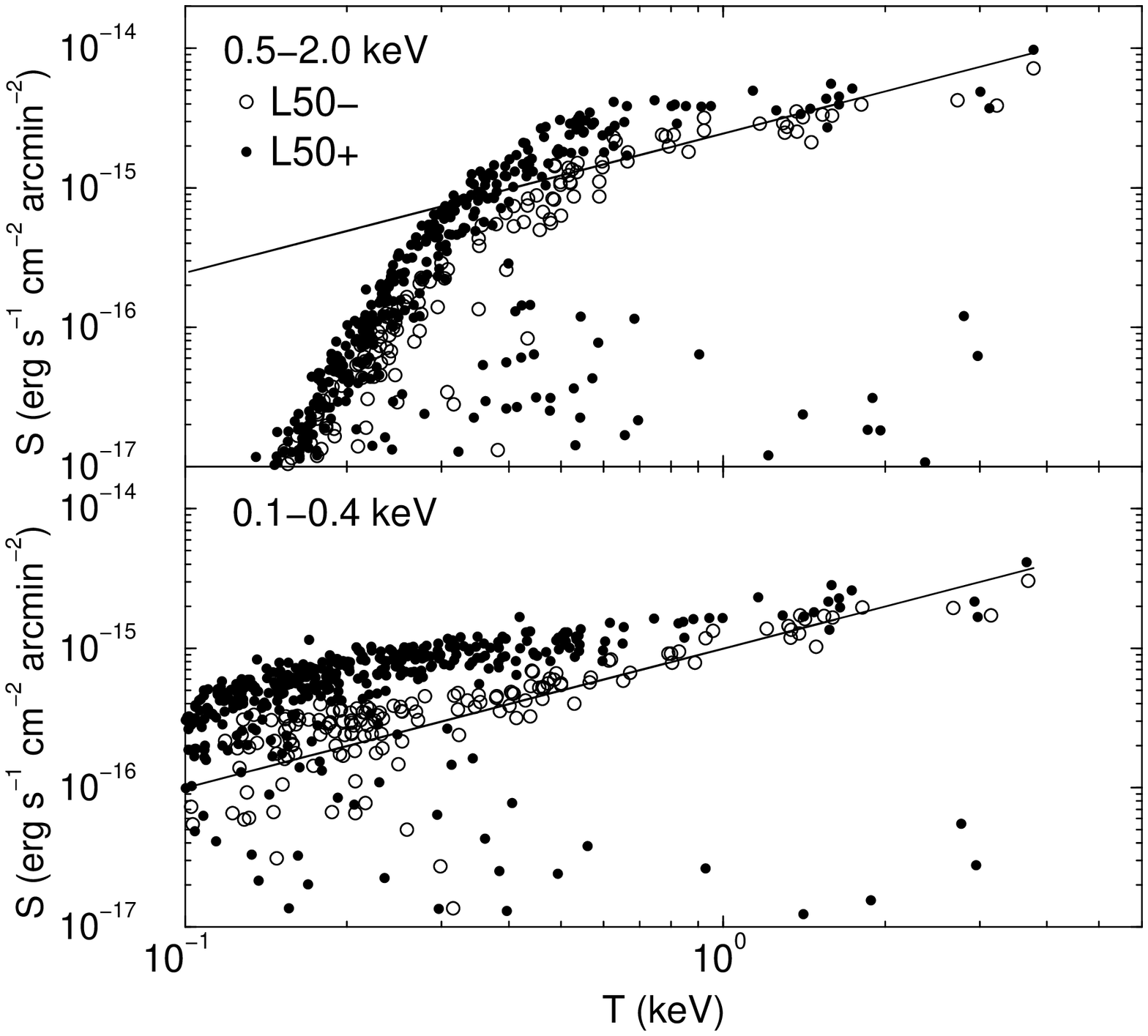}}
\figcaption{The mean surface brightness plotted against
(luminosity-weighted) temperature for groups and clusters identified
in the L50+ and L50- simulation at $z=0.4$. The top panel shows the
relation for the 0.5-2.0 keV band while the bottom shows the 0.1-0.4
keV band.  Also plotted is the power-law relation based on scaling
arguments and free-free emission (see text).
\label{fig:sb_tlum}}
\vspace{\baselineskip}

The figure shows more clearly the effect of resolution: while the
largest clusters have the same predicted mean surface brightness in
both simulations (which vary only in resolution), the smaller groups
have systematically underpredicted emission.  We can also check the
surface-brightness-temperature ($S-T$) relation against simple
analytic predictions.  If we assume that all collapsed objects have
the same profile when density is scaled by the critical
density of the universe and radius is scaled by the virial radius, then
the bolometric free-free luminosity should vary approximately as $L
\sim T^2 H$ (see also Bryan \& Norman 1998; Voit 2000).  Combining
this with $r_{200} \sim T^{1/2} H^{-1}$, we obtain:
\begin{equation}
S \sim T H^3 (1+z)^{-4}.
\end{equation}
Note the relatively slow evolution with redshift; for $\Omega = 1$,
this becomes $S \sim T (1+z)^{1/2}$.  This line is plotted in
Figure~\ref{fig:sb_tlum}.  The well-resolved clusters climb above this
line around 1 keV due to the increased importance of line-emission,
which the simple analytic estimate does not include.  Also, clusters
with temperatures below about 0.2 keV produce very few photons in the
0.5-2.0 keV band.  The points with low surface-brightness but high
temperature represent relatively small halos which are in the
processes of merging with larger systems.

\vspace{\baselineskip}
\epsfxsize=3.5in
\centerline{\epsfbox{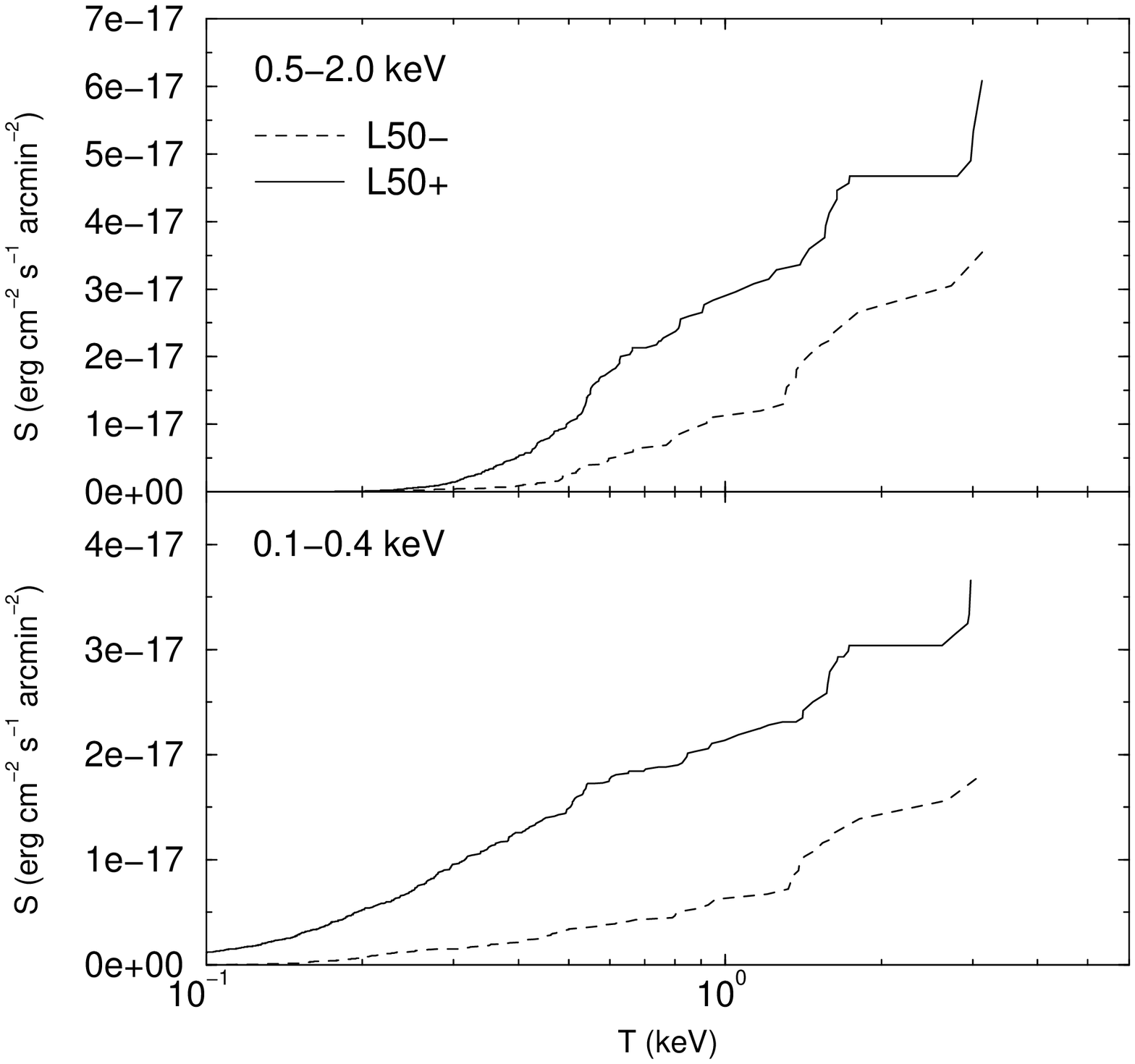}}
\figcaption{The cumulative contribution to the mean surface brightness
from the clusters and groups identified in the same simulation volume
discussed in the previous two figures. As usual, the top (bottom)
panel shows the 0.5-2.0 (0.1-0.4) X-ray band.  Within each figure, the
solid and dashed lines show the effect of numerical resolution.
\label{fig:sb_cum}}
\vspace{\baselineskip}

While useful, this figure still doesn't fully answer the question of
what temperature range of halos is responsible for the X-ray
background.  This is addressed in Figure~\ref{fig:sb_cum}, which shows
the cumulative contribution as a function of cluster or group
temperature.  Although a wide range of objects contribute, it is clear
that the dominant contribution comes from halos with temperatures of
order 1 keV ($\sim 10^{14} M_{\odot}$).  For example, in the harder
band, 50\% of the flux is contributed by halos with temperatures
greater than 1.0 keV, while for the softer band, a similar fraction
comes from groups with temperatures greater than 0.6 keV.

In each case, the total flux from identified halos is within a few
percent of the flux as computed by summing all pixels of the
surface-brightness map, indicating that filaments contribute a
negligible fraction of the total flux.  We should also point out that
a substantial fraction of the background comes from relatively large,
2-4 keV objects, which are mostly missing in the smallest L20
simulations described in Table~\ref{table:sims}.  This helps to
explain why those simulations have a systematically different 0.5-2.0
keV surface brightness distribution function.


\subsection{Comparison to observations}
\label{sec:observations}

We can compare the predicted mean background fluxes from
Equations~\ref{eq:slim_0.5_2.0} and \ref{eq:slim_0.1_0.4} to those
determined observationally.  At 1 keV, the background has an intensity
of about $I_X = 10$ \sbspunits with a spectral slope of $\alpha
\approx 1.0$ (e.g. Wu \etal 1991; Gendrau \etal 1995; Barcons, Mateos
\& Ceballos 2000).  At least 70\% of this has been resolved into point
sources (Hasinger \etal 1998; Giacconi \etal 2000) and so does not
originate in the diffuse gas of interest here.  Therefore, we can
convert this into an upper limit on the observed diffuse background;
in the 0.5-2.0 keV band, this becomes:
\begin{equation}
S^{obs}_{0.5-2.0} < 5.0 \times 10^{-16} \hbox{\sbunits}.
\label{eq:sobs_0.5_2.0}
\end{equation}

From a review of shadowing experiments conducted by ROSAT, Warwick \&
Roberts (1998) found a mean background intensity in the 0.1-0.4 keV
band (i.e. at 0.25 keV) of 20-35 \sbspunits, and that at least 80\% of
this was resolved into background sources.  Assuming a spectral slope
of $\alpha = 2$ (e.g. Gendreau \etal 1995; Barber, Roberts \& Warwick
1996).
\begin{equation}
S^{obs}_{0.1-0.4} < 3.0 \times 10^{-16} \hbox{\sbunits}.
\label{eq:sobs_0.1_0.4}
\end{equation}
A reasonable extrapolation of AGN properties indicates that the likely
upper limit to the diffuse background in this band is 4 \sbspunits
(Wu, Fabian \& Nulsen 2000b), or $1.9 \times 10^{-16}$ \sbunits.

Comparing these observations to the predicted values, it's clear that
unless $\Omega_b h^2$ is far lower than the value we expect, the
predicted background is larger than that observed.  Even for the case
with 1.5 keV of feedback, we would have to reduce the baryon fraction
by 60\%, to $\Omega_b h^2 = 0.011$, considerably below essentially all
estimates (Burles \etal 1999).  Reducing the assumed metallicity of
the gas ($Z=0.3$ of solar) would trim the predicted background
somewhat, but since most of the flux comes from objects in the 1-3 keV
range, where the metallicity is relatively well measured, there does
not appear to be much room for maneuver.

In order to get an idea of how much feedback would be required to
match observations, we performed two additional experiments.  In one,
we added the energy at $z=1$ instead of $z=3$ as would be required in
order not to violate the IGM temperature constraints from Ly$\alpha$
clouds (e.g. McDonald \etal 2000; Schaye \etal 1999; Bryan \& Machacek
2000).  This actually resulted in an increased diffuse background,
mostly because the effect is larger when energy is added at lower
densities, before the clusters and groups have fully formed.  We
speculate that adding energy earlier ($z > 3$) would be more
efficient; however, it's not obvious what physical mechanism could
provide so much energy so early.

We also examined the effect of increasing the feedback energy to 4.5
keV instead of 1.5 keV (at $z=3$).  For the low-resolution simulation
(L50-), this resulted in a soft background of $1.9 \times 10^{-16}$
\sbunits, somewhat below the most conservative observed limit quoted
above.  Since we expect about of factor of two increase in going to
the L50+ resolution level (which would push the predicted background
above even the conservative limit), it seems likely that a substantial
amount of heating (at least 5 keV per particle) will be required.
It should be stressed that this energy input occurs everywhere in our
model, but it is certainly possible and even likely that energy would
be more efficiently liberated (by star formation for example) in
high-density regions than low-density regions.  This would decrease
the total energy budget required.


\subsection{Comparison to previous work}

The first paper to make specific predictions about the distribution of
the soft X-ray background was an ambitious work by Scaramella, Cen \&
Ostriker (1993) who used numerical simulations to generate artificial
maps of the X-ray background at 1 and 2 keV (as well as the
Sunyaev-Zel'dovich $y$ parameter).  Because of numerical limitations,
they were forced to use a more complicated chaining method of the
boxes as well as being unable to check the convergence properties of
the result.  Their predicted mean background intensity was $I_X =
0.02$ \sbspunits at 1 keV, much lower than found here.  Although it is
possible that the different cosmological model (a closed $\Omega=1$
CDM variant) played a role, there are two significant differences.
One is that they removed the brightest pixels under the assumption
that they would not be counted by observers; the second stems from the
substantially lower resolution available at the time.  Their spatial
resolution was typically five times worse than used here and given the
sensitivity on resolution we have highlighted earlier, this might
explain the discrepancy.  Still, they found a profile at the high end,
$S dP/dS \sim S^{-1.8}$, which adequately matches the appropriate part
of our distribution.  The reason for this is given in Voit \& Bryan
(2001).

The mean background from diffuse gas (although not the distribution
function itself) was calculated in an approximate fashion by Pen
(1999).  He used numerical simulations to determine the mean clumping
factor of the IGM and then adopted a mean temperature from the cosmic
virial theorem to generate an estimate of the mean background,
assuming a mean metallicity of 0.25 solar.  He found that numerical
simulations gave only a lower limit to the mean clumping factor but
that this lower limit was substantially larger than implied by the
observed background.  This is in agreement with the results found
here, that the predicted background is approximately an order of
magnitude larger than observed, indicating the need for some sort of
additional physics.  Pen also estimated that if feedback is
responsible, the required energy budget would be around 2 keV, a value
somewhat smaller than derived here.

Wu, Fabian \& Nulsen (2000b) employed a semi-analytic technique to
compute the predicted amplitude and spectrum of the diffuse X-ray
emission.  Although most of their models included cooling, a model
without cooling or stellar feedback produced a mean background level
which is substantially in agreement with that found here.  They argued
that while cooling would help somewhat, additional heating would be
required and derived an excess specific energy of about 1 keV per
particle.  This is substantially lower than what we find in this
paper; however, Wu, Fabian \& Nulsen also included a treatment of
cooling, making the comparison more difficult.

More recently, Dav\'e \etal (2000) looked at the distribution of the
warm-hot intergalactic medium (WHIM) and without performing a full
calculation speculated that the WHIM gas would not overproduce the
X-ray background.  Here, we have shown that without additional heating
or cooling this is not the case (although we should note that most of
the simulations analyzed in that paper did model cooling and stellar
feedback).

Finally, as this paper was in the late stages of preparation, two
preprints appeared (Croft \etal 2000; Phillips, Ostriker \& Cen 2000)
using simulations to perform a full calculation of the X-ray
background similar to that done here.  The two paper use different
simulation methods but similar physical models (radiative cooling and
a simple feedback).  They both concluded that the predicted background
from diffuse gas is within the observed bounds, a result which at
first appears to contradict this paper.  There are several possible
resolutions to this apparent conflict.  

The first we examine is resolution and box size: in this paper we have
carefully controlled for both of these effects, demonstrating that
although a relatively modest simulation volume will produce a good
estimate of the background (at least for the softer bands), numerical
resolution is very important.  Philips, Ostriker \& Cen (2000) have
good mass resolution but relatively poor spatial resolution: they use
a dark matter particle mass of $9 \times 10^9 M_{\odot}$, but a fixed
cell size of only 195 $h^{-1}$ kpc.  A direct comparison is difficult
since the algorithm differs from that used here; however, experience
with such simulations (Bryan \& Norman 1998) indicates that this cell
size would lead to a significant underprediction of the luminosity for
a simulation without cooling or radiative feedback.  However, it is
possible that this resolution is sufficient when these two processes
are also included.  Croft \etal (200), using a smoothed-particle
hydrodynamics technique, have both high spatial and mass resolution (7
$h^{-1}$ kpc minimum smoothing length and $7\times 10^9 M_{\odot}$
dark matter particle mass).  According to the results presented here,
this should be sufficient resolution to obtain a reasonable estimate
of the diffuse X-ray background.  This, combined with the fact that
the two groups produced similar results with different techniques,
makes it unlikely that resolution and box size are playing a large
role.

However, it is important to recognize that both Croft \etal and
Philips \etal included radiative cooling and feedback.  Since the
cosmological models were quite similar and previous comparisons
between the different simulation techniques have produced similar
results (Dav\'e \etal 2000), the difference is probably due either to
cooling or feedback.
Although there are still many uncertainties, it appears unlikely that
feedback is the culprit since --- as Croft \etal note --- in SPH
simulations the energy is liberated in dense regions and quickly
radiated away.  This would be consistent with the (unrealistically?)
large levels of feedback required in this paper to bring the predicted
background down to the level of the observation limits.

This leaves cooling.  Because of numerical difficulties in treating
the steep density profiles that radiative cooling generates, Croft
\etal used a post-processing step to separate the cold phase from the
hot phase, substantially decreasing the density of the hot phase.
This diminished the emission by a factor of 15, so clearly the
treatment of cooling is extremely important.  A prominent role for
cooling has also recently been proposed as a explanation for the
scaling properties of clusters and groups (Bryan 2000).  In that work,
it was suggested that the varying efficiency of cooling in groups and
clusters creates an effective entropy floor (as is also observed).
Although we cannot directly test the cooling hypothesis in this work,
we can examine the effect of adopting an entropy floor.  To test this,
we set the density of the gas in virialized regions (defined as those
regions for which the density is at least 200 times that of the mean
density), such that the ``entropy'' is increased by a value $k T /
n_e^{2/3}$ = 100 keV cm$^{2}$, where $n_e$ is the electron density and
$T$ the temperature.  If this is done, then the resulting X-ray
background prediction in the 0.5-2.0 keV band drops below observed
limits, although the 0.1-0.4 keV band is still marginal (see 
Voit \& Bryan 2000 for more details).


\section{Conclusions}

In this paper, we have used numerical simulations of a popular
$\Lambda$-dominated cosmology to predict the contribution of the hot
diffuse gas to the X-ray background.  We assumed a constant
metallicity of 0.3 solar and examined two bands: 0.1-0.4 keV and
0.5-2.0 keV, computing the mean background as well as the surface
bright distribution function.  The simulations included dark matter
and baryons but did not self-consistently model either radiative
cooling or star formation.  We did however, examine the general effect
of energy injection due to star formation via a simplified model of
uniform heating.  In order to focus on the diffuse gas, we ignored the
contribution from compact sources.  The conclusions are as follows:
\begin{itemize}
\item The predicted mean diffuse background depends on the mass
resolution of the simulation, systematically increasing as the
resolution improved.  Although we feel we are nearing numerical
convergence, this implies that our results are lower limits to the
mean background.
\item The surface brightness distribution $(dP/dS)$ in the 0.5-2.0 keV
band is only weakly dependent on resolution, and that at the very
highest surface brightness end (unfortunately, it is also this end
which makes the primary contribution to the mean background).  The
softer 0.1-0.4 keV band is less well resolved and the entire
distribution shifts with resolution.
\item Smoothing causes the distribution function to peak at moderate
surface brightness values.  While slight smoothing (of only a few
arcsecs) has only a small effect, even 10'' or 30'' smoothing
significantly modifies the shape of the distribution function.
\item Although a substantial fraction of the baryonic mass is in the
form of filaments or other low-density structures, the diffuse
background overwhelming originates in groups and clusters.  The
0.1-0.4 keV band comes from groups with a virial temperature of order
0.6 keV, while the harder 0.5-2.0 keV band originates in larger, 1.0
keV systems.  However, in both cases, a wide range of objects
contribute.
\item Feedback modifies the shape of the distribution function.
This occurs in large part because the additional
energy inflates the core of clusters and groups,
decreasing the emission from the brightest central
regions and thereby suppressing the high end of
the distribution function.
\item Since the measured X-ray background has now been largely
resolved into point sources, presumably almost entirely due to compact
sources, this provides an upper limit to any possible contribution
from diffuse gas.  Our upper limit to the predicted mean distribution
substantially exceeds this bound.  This confirms the initial
suggestions (Pen 1999; Wu, Fabian \& Nulsen 2000b) that the hot gas in
clusters and groups --- in the absence of non-gravitational heating or
radiative cooling --- would overproduce the X-ray background.
\item Including a simple form of feedback (a uniform 1.5 keV per
particle injected at $z=3$) reduces but does not eliminate the
discrepancy.  It seems likely that either (a) much more energy is
required (at least 5 keV per particle), (b) the energy is injected at
substantially higher redshift than three, or (c) additional physical
processes are at play.  This could be a more complicated form of
feedback which operates in a substantially different fashion than
simple energy injection, or it could be that a substantial fraction of
the gas in groups cools, reducing the luminosity of such groups (Bryan
2000).  If this latter suggestion is correct, it would be consistent
with the differences between our results and another recent simulation
(Croft \etal 2000) which included radiative cooling but did not exceed
the observational limits.
\end{itemize}

While this work is a significant step forward in the study of the
diffuse X-ray background, much more work remains to be done.  This
includes understanding the effect of realistic energy and metal
injection on the IGM.  It will be very important to understand the
role of cooling on the state of the hot gas in groups and clusters.

The X-ray background contains a wealth of potential information
relating to the diffuse, cosmologically distributed gas that
hierarchical cosmological models predict.  This information is
difficult to interpret since it requires removing the contribution
from point sources and accounting for the effect of the lower number
of photons coming from areas of low intrinsic surface brightness.
However, a careful measurement of the surface brightness distribution
function would constrain the thermal history of the gas, helping us to
understand the energy input from supernovae and AGN.

\acknowledgements
                                                                       
Support for GLB was provided by NASA through Hubble Fellowship grant
HF-01104.01-98A from the Space Telescope Science Institute, which is
operated by the Association of Universities for Research in Astronomy,
Inc., under NASA contract NAS6-26555.  Some simulations in this paper
were performed at the National Center for Supercomputing Applications
(NCSA).


\end{document}